\documentclass[12pt,preprint]{aastex}

\def\lea{\mathrel{<\kern-1.0em\lower0.9ex\hbox{$\sim$}}}
\def\gea{\mathrel{>\kern-1.0em\lower0.9ex\hbox{$\sim$}}}

\shorttitle{On Identifying the Progenitors of Type Ia Supernovae}
\author{
Mario Livio\altaffilmark{1} and 
J.~E.\ Pringle\altaffilmark{1,2} 
}

\shortauthors{Livio and Pringle}
\email{mlivio@stsci.edu}
\altaffiltext{1}{Space Telescope Science Institute, Baltimore, MD, USA}
\altaffiltext{2}{Institute of Astronomy, Cambridge University, Madingley Road, Cambridge CB3 0HA, UK}

\begin{document}

\title{On Identifying the Progenitors of Type Ia Supernovae}

\begin{abstract}
We propose two new means of identifying the main class of progenitors of Type~Ia supernovae---single or double degenerate: (i)~If the range of supernova properties is significantly determined by the range of viewing angles of non-spherically symmetric explosions, then the nature of the correlation between polarization and another property (for example, the velocity gradient) can be used to determine the geometry of the asymmetry and hence the nature of the progenitor, and (ii)~in the double- but not in the single-degenerate case, the \textit{range} in the observed properties (e.g., velocity gradients) is likely to increase with the amount of carbon seen in the ejecta.
\end{abstract}


\section{INTRODUCTION}
The fact that we are still uncertain about the nature of the progenitor systems of Type~Ia supernovae (SN\,Ia)---some of the most powerful explosions in the universe---has become a major embarrassment for modern astrophysics. The problem is compounded by the fact that Type~Ia supernovae (SN\,Ia) have become a key tool for placing constraints on the nature of the dark energy that propels the accelerating cosmic expansion (Riess et~al.\ 1998; Perlmutter et~al.\ 1999). The use of SN\,Ia for the determination of such cosmological parameters requires the taming of evolutionary effects---an understanding, or at least a calibration, of the evolution of SN\,Ia luminosity with cosmic time. A full understanding of these effects must hinge on a proper comprehension of the nature and properties of the progenitors (see, e.g., Livio 2000; Nomoto et~al.\ 2000; Hillebrandt \& Niemeyer 2000 for reviews, and Riess \& Livio 2006 for a more recent discussion).

In the present very brief letter we present two simple \textit{ideas} for observations and analyses that, while not easy, could be used to identify the dominant (in terms of its contribution to the SN\,Ia rate) class of progenitors.

\section{THE TWO POTENTIAL PROGENITOR SYSTEMS}
There is little doubt that most SN\,Ia represent thermonuclear disruptions of mass-accreting Carbon-Oxygen (CO) white dwarfs (WDs), when these white dwarfs reach the Chandrasekhar limit and ignite carbon at their centers (see, e.g., Livio 2000, for a detailed discussion).  This model is generally accepted even though many of the details of the process are not fully understood, with one major uncertainty concerning the flame propagation itself, and in particular the transition from subsonic deflagration to supersonic detonation (see, e.g., H\"oflich et~al.\ 2010 for a recent discussion).

There are two scenarios that have been traditionally identified for the progenitor systems (and discussed for almost four decades). 

\subsection{Single-degenerate Scenario (SD)}

In the single-degenerate (SD) scenario, a CO WD accretes hydrogen-rich or helium-rich material from a main-sequence, subgiant, or giant companion at a  steady rate that allows for the WD to grow in mass (e.g., Whelen \& Iben 1973; Nomoto 1982). The explosion occurs when carbon ignites, as the mass of the accreting white dwarf approaches the Chandrasekhar limit. Potential progenitor systems of this type include, for instance, the supersoft x-ray sources and recurrent novae, assuming that the WDs in these systems are not of O-Ne composition. One of the main strengths of the SD scenario is the fact that, in principle at least, one could get two populations in terms of time delays (as appears to be observed, e.g., Howel et~al.\ 2007): relatively prompt explosions ($t_\mathrm{delay} \lea 200$~Myr) could result from the WDs accreting from young main-sequence companions, while late ones ($t_\mathrm{delay}\sim2$--3~Gyr) from the WDs accreting from low-mass red giants. The fact that it is apparently very difficult to detect the hydrogen from the non-degenerate component remains a major stumbling block for a definitive confirmation of the SD scenario. 

\subsection{Double-degenerate Scenario (DD)}

In the double-degenerate (DD) scenario, two CO WDs in a binary system are brought together by angular momentum and energy loss caused by the emission of gravitational radiation and coalesce (e.g., Webblink 1984; Iben \& Tutukov 1984). Progenitor systems of this type are binary WD systems (in which the total mass exceeds the Chandrasekhar limit) with binary periods shorter than $\sim$13 hours (to allow merger within a Hubble time).  At the same time, we should note that the DD scenario quite naturally reproduces what appears to be the delay-time distribution (DTD) of: DTD~$\sim t_\mathrm{delay}^{-1}$ (e.g., Maoz et~al.\ 2010, 2011).  The key problem with the DD scenario is a theoretical one. In the merger process, carbon is expected to ignite near the surface (e.g., Mochkovitch \& Livio 1990), transforming the C-O into an O-Ne mixture (e.g., Kawai, Saio \& Nomoto 1987), which is expected to lead to collapse rather than to explosion (e.g., Nomoto \& Kondo 1991; Canal 1997). While a few recent theoretical works suggest that this problem may not be altogether insurmountable (e.g., Pakmor et~al.\ 2010; Yoon, Podsiadlowski \& Rosswog 2007; Rosswog et~al.\ 2009), a detailed three-dimensional calculation of all the processes involved is badly needed.

\section{ASYMMETRIES AND DISTINGUISHING OBSERVATIONS}
The ambiguous situation concerning the nature of the progenitors clearly calls for theoretical predictions that can be tested observationally. An excellent,  detailed attempt in this direction has recently been published by Blondin et~al.\ (2011).  In another recent work, Sternberg et~al.\ (2011) have shown evidence for gas outflows from some progenitor systems of SN\,Ia in nearby spirals. These findings are consistent with the SD scenario (but see also Badenes et~al.\ 2007). In this section we present two new \textit{ideas} for distinguishing characteristics.  These ideas were broadly inspired by the history of gamma-ray bursts (GRBs), where it turned out that identifying a particular two-dimensional phase space (hardness-duration; e.g., Kouveliotou et~al.\ 1993; Dezalay et~al.\ 1995) revealed the presence of two classes of progenitors.

\subsection{Explosion Asymmetry}

If all SN\,Ia explode in a spherically symmetric manner, then it would be difficult to distinguish between the possible progenitors. However, there is evidence that the explosions are not spherically symmetric. We do note, nonetheless, that while there have been some claims for evidence for asymmetries in SN\,Ia (e.g., Maeda et~al.\ 2010; Maund et~al.\ 2010;Wang \& Wheeler 2008), the more extensive study by Blondin et~al.\ (2011) has shown that at least some anisotropic configurations probably do not occur in nature.  Here we start by characterizing asymmetries in a simple manner, depending on the nature of the progenitor.

\noindent (i) Single Degenerates

If the progenitor is a SD, then the most likely source of potential asymmetry (if one exists at all) is a somewhat off-center ignition of the explosion. For example, Maeda et~al.\ (2010) state that ``ignition at an offset from the center is a general feature of Type~Ia supernovae." They argue that the variations seen in SN\,Ia can be accounted for by viewing the off-center explosion of a single degenerate progenitor from different angles (see also Benetti et~al.\ 2005). To characterize this possibility, and to keep it simple, we may think of the resulting explosion as being broadly axially symmetric, so that the deviation from spherical symmetry corresponds predominantly to an \boldmath{$m=1$} mode.  That is, the observed character of the explosion would be observed to vary in a monotonic manner \textit{as the viewing angle moves from the WD's north pole to its south pole}. Maeda et~al.\ (2010) give an example of a numerical model in which if the explosion is viewed from one pole it is seen as a high-velocity gradient (HVG) event  at early phases with red shifts in late-time emission lines, whereas if it is viewed from the other pole it is seen as a low-velocity gradient (LVG) event with blue shifts at late phases.

\noindent (ii) Double Degenerates

In the double-degenerate case, the less massive WD fills its Roche lobe first. The resultant mass transfer is dynamically unstable and this results in the lower mass companion being totally dissipated within a few orbits to form a thick disk around the primary (e.g., Benz et~al.\ 1990; Mochkovitch \& Livio 1989; Segretain, Chabrier \& Mochkovitch 1997). Assuming that accretion from this disk can indeed produce an explosion (see the difficulty with ignition near the surface mentioned in the previous section), with ignition occurring close to the center of the more massive degenerate star, then the most likely source of asymmetry results from the fact that when the explosion occurs, the primary WD is still surrounded by a massive belt of effectively unaccreted (and unburned) material.  In this case, the explosion may be expected to be axially symmetric, but predominantly of the form \boldmath{$m=2$}. That is, there is an additional reflective symmetry about the equator. Here the character of the explosion changes monotonically \textit{as the viewing angle moves from the pole to the equator}. In contrast to the SD case (e.g., Maeda et~al.\ 2010), detailed numerical simulations of such explosions have yet to be carried out. 

\subsection{Polarization}

The difference in the symmetry properties of the two scenarios suggests that one could (in principle at least) use polarization measurements taken at \textit{early times} (prior to maximum light) to distinguish between them.  An excellent summary of the polarization properties of supernovae is given by Wang \& Wheeler (2008). They note that Type~Ia SN display modest continuum polarization, but strong line polarization, clearly observed only prior to maximum light, and that the polarization declines after maximum (in particular in the high-velocity Ca\,II IR triplet). Maund et~al.\ (2011) interpret this as meaning that the ioniza\-tion/density structure becomes more spherically symmetric with depth. Maund et~al.\ (2010) note that simple models seem to indicate that departures from spherical symmetry are less than of order 10 percent, and that correlations have been found between polarization and observed explosion parameters. A detailed demonstration of how difficult it is to determine the intrinsic polarization is given by Howell et~al.\ (2001). These authors consider only idealized models with oblate spheroidal asymmetry and conclude that both the flux and polarization of SN~1999by can be reasonably well produced by models with asphericity of around 20 percent.

\begin{figure}
\plotone{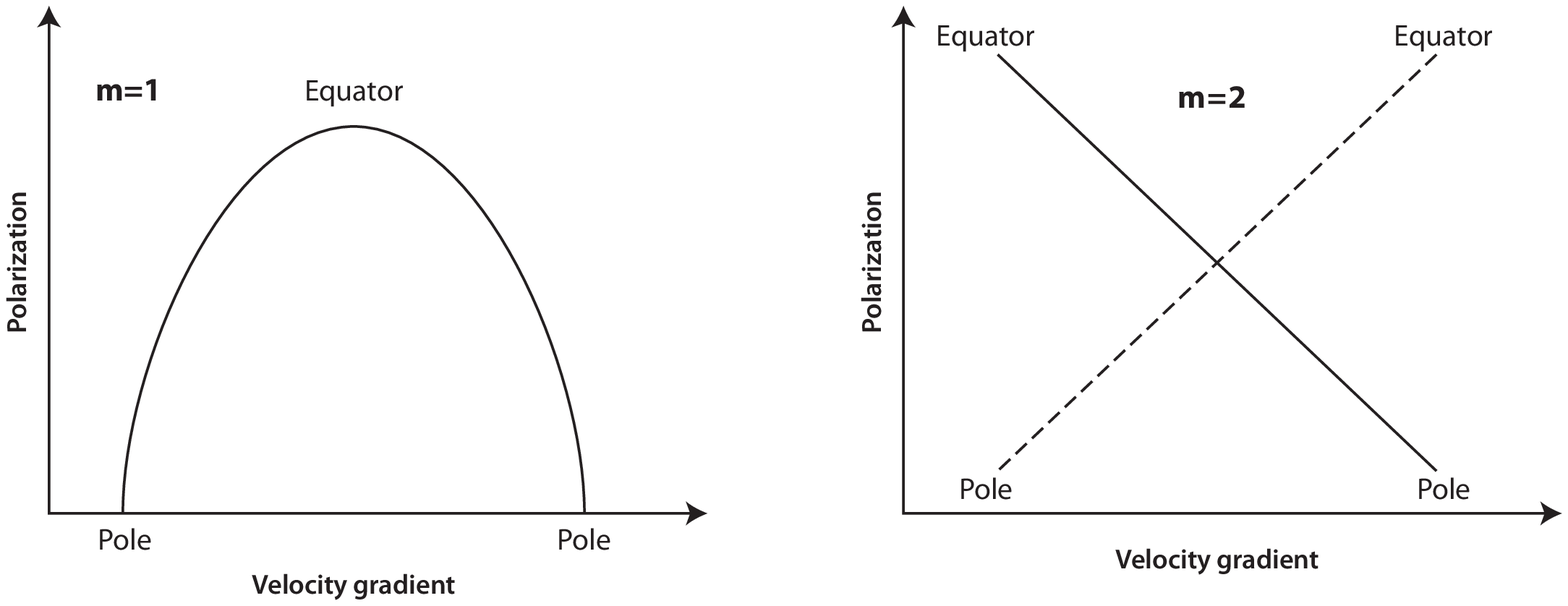}
\caption{A schematic representation of the expected polarization as a function of the velocity gradient for the SD scenario in the case of off-center ignition (left), and for the DD scenario (right). In the SD case, we expect the velocity gradient to be a two-valued function of polarization, with largest and smallest values corresponding to essentially zero polarization. In the DD case we expect observed SN properties (here velocity gradient) to be a single-valued and monotonic function of polarization. See text.\label{fig1}}
\end{figure}

What we note here is that \textit{the nature of the correlation between polarization and observed SN properties} can be used to distinguish between the two scenarios. As a specific example we consider possible correlations between the  polarization and the velocity gradient. We give a schematic representation of such possible correlations in Figure~\ref{fig1}. Because of the axial symmetry, the SD model predicts zero polarization when viewed from each pole. From simple geometrical considerations (e.g., Maeda et~al.\ 2010)  one would then expect \textit{both} the highest \textit{and} the lowest velocity gradients to correspond to viewing angles at the individual poles, with maximum polarization occurring for intermediate values of the velocity gradient.  Thus in the SD case we might expect the observed velocity gradient to be a \textit{two-valued} function of the observed polarization. In the DD model, high polarization still corresponds to viewing from the equator, and (by symmetry) zero polarization to viewing from either pole. But in this case, the velocity gradient  would be \textit{the same} as viewed from either pole, and would differ most when viewed from the equator. Detailed models will be required to determine whether the velocity gradient is larger or smaller when the explosion is viewed from the equator---it might be that the velocities are reduced by the overlying ring of unaccreted material, or that the higher opacity of the overlying material means that the photospheric velocity is higher. In either case, for the DD scenario, we might expect the relationship between velocity gradient and polarization to be \textit{single-valued} and monotonic. Schematically, the results would appear as in Figure~\ref{fig1}.

\subsection{Unburnt Fuel}
\begin{figure}
\plotone{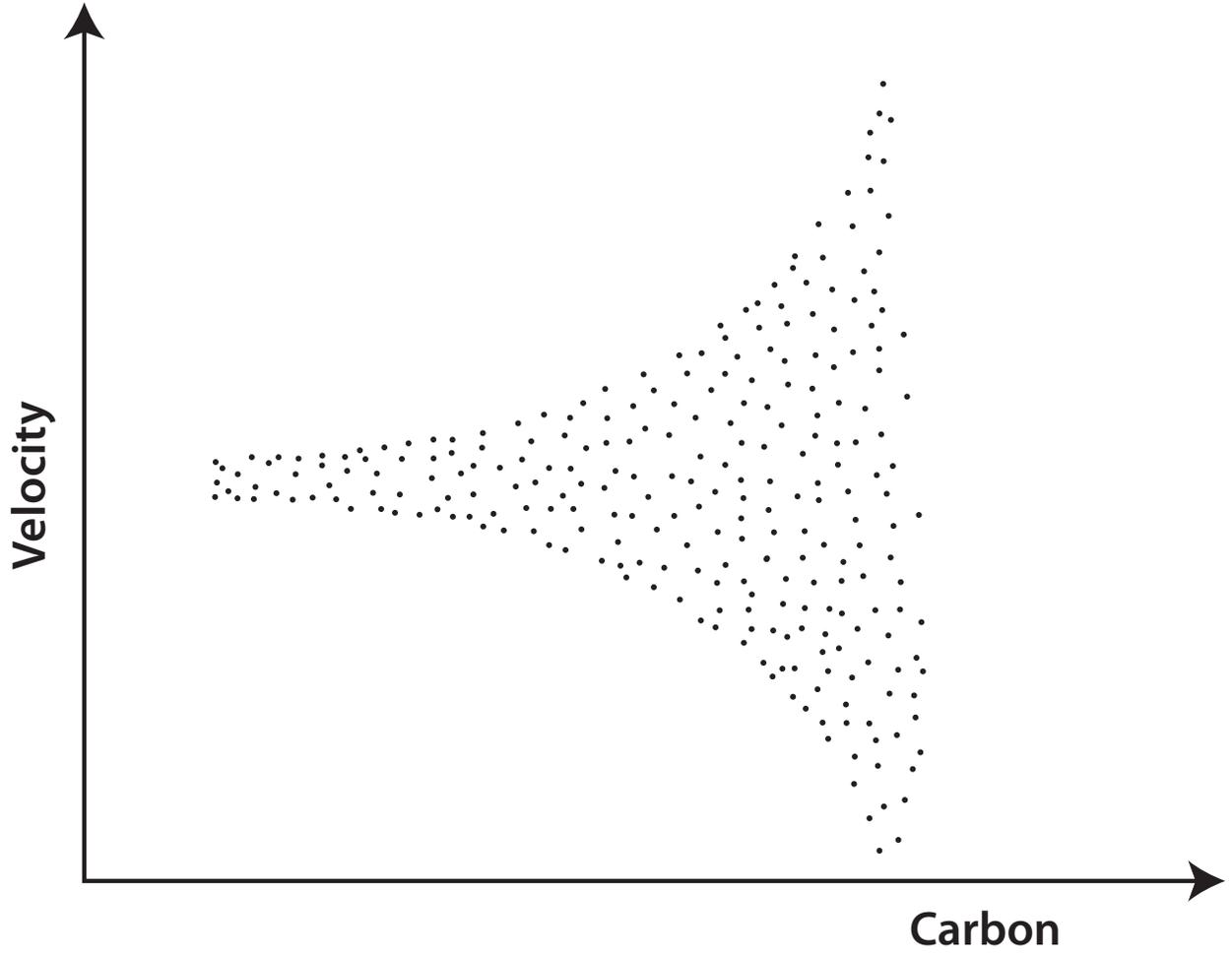}
\caption{A schematic representation of the expected \textit{range} in velocities as a function of the observed amount of carbon, in the DD scenario. No such correlation is expected for the SD scenario. See text.\label{fig2}}
\end{figure}

We have commented above that one distinguishing feature of the SD scenario is that the ejecta should contain unburnt fuel originating in the outer layers of the companion star. Detection of hydrogen among the ejecta would clearly give a definite boost to the likelihood that the progenitor was a single degenerate, but this would require an improvement by about two orders of magnitude in the sensitivity (see e.g., Livio 2000 for a discussion). In the DD scenario the primary white dwarf is likely to be surrounded by a massive, equatorial torus of carbon and oxygen when the explosion ignites. The size, distribution and elemental make-up of the torus must depend primarily on the mass of the original secondary white dwarf. Most of the torus is likely to remain unburnt during the nuclear explosion, but is likely to form part of the supernova ejecta.  The more massive the torus, the more substantial is likely to be the effect on the velocity distribution of the ejecta and on the apparent elemental abundances. Determining the effects from a quantitative standpoint will require detailed numerical simulations. As with hydrogen, measuring the mass of the amount of unburnt carbon in the ejecta is not easy but some progress is being made (e.g., Parrent et~al.\ 2011; Taubenberger et~al.\ 2011). A large, unbiased sample of pre-maximum spectra will be needed, coupled with a detailed analysis of the carbon features (e.g., C\,II~$\lambda$6580 and $\lambda$7234).  What we note here is that, given the range of viewing angles, we might expect that the \textit{range} in the observed velocities (and perhaps even luminosities) is likely to increase with increasing amount of carbon observed.  This is depicted schematically in Figure~\ref{fig2}.  No such correlation between the \textit{range} of velocities and the amount of carbon would be expected for the SD scenario. 

\section{CONCLUSIONS}
We have proposed two new ideas for observations that could potentially identify the main class of progenitors of SN\,Ia---either single degenerate or double degenerate. Clearly, to allow for a detailed comparison between theory and observations, full-scale 3D simulations that include all the relevant processes will be required.  However, the results of such simulations will also necessarily depend on the input parameters and physics. Here we believe that we have identified robust trends, that can be used as distinguishing characteristics  between the two scenarios.  As excellent observations of velocity gradients are becoming available (e.g., Foley, Sanders \& Kirshner 2011), coupling those with polarization measurements and measurements of the amount of carbon should become feasible in the not too distant future.

\begin{acknowledgements}
We acknowledge helpful discussions with Louis Strolger, Adam Riess, and Andy Howell.  JEP thanks the Distinguished Visitor Program at STScI for its continued hospitality.  We thank the referee for his/her helpful comments.
\end{acknowledgements}

\clearpage

\end{document}